\documentclass{article}
\usepackage[english]{babel}
\usepackage{graphicx}
\usepackage{subcaption}
\usepackage[utf8]{inputenc}
\usepackage[export]{adjustbox}
\usepackage{wrapfig}
\usepackage{authblk}
\graphicspath{ {./images/} }
\title{A System Dynamics Model of Bitcoin: \\ Mining as an Efficient Market\\ and the Possibility of ‘Peak Hash’}

\author[1]{Davide Lasi (lasi@mit.edu)}
\author[2]{Lukas Saul (saul@hfut.edu.cn)}
\affil[1]{System Design \& Management (SDM) Fellow, MIT}
\affil[2]{Department of Electrical Science, Hefei University of Technology}

\date{18 April 2020}

\begin{document}

\maketitle

\begin{abstract}
The mining of bitcoin is modeled using system dynamics, showing that the past evolution of the network hash rate can be explained to a large extent by an efficient market hypothesis applied to the mining of blocks. The possibility of a decrease in the network hash rate from the next halving event (May 2020) is exposed, implying that the network may be close to 'peak hash', if the price of bitcoin and the revenues from transaction fees will remain at approximately the present level.
\end{abstract}

\section{Introduction}

Bitcoin is the first publicly-issued cryptocurrency, introduced in 2008 by Satoshi Nakamoto with a paper \cite{satoshi} that discloses not only the nuts and bolts of the underlying technology, the blockchain, but also the expected dynamics of the complex system created by the adoption of this new form of electronic money. These dynamics are crucial because they are what make the bitcoin and other cryptocurrencies robust, scalable, and secure. This paper aims to shed some light on one of these dynamics, which is at the core of bitcoin’s security: the mining of blocks.

By applying a modeling technique known as system dynamics \cite{forrester, sterman}, the evolution of the network hash rate is shown to be explainable, to a large extent, with a simple efficient-market hypothesis, where bitcoin miners mine because of the expected profit they make from mining, net of the cost of electricity. On the one hand, this result is unimpressive and reassuring, in that it resonates well with common sense. On the other hand, it may have significant implications on the future evolution of the network hash rate.

In particular, if the profit from mining becomes negative as a consequence of a price crash or of a dramatic fall in the reward, the network hash rate might drop significantly, because hashing power would be diverted to the mining of more profitable alternative coins. This might be the case, for instance, in the aftermath of the forthcoming halving of May, 2020, when the network could have reached the ‘peak hash’. Possible consequences include higher volatility of the bitcoin network hash rate, as well as security issues.

\section{The Bitcoin and System Dynamics}

System dynamics (SD) is an approach to the modeling of systems in terms of stocks, flows, and feedback loops, introduced in the 60s by MIT’s Prof. Jay Forrester \cite{forrester}. It became very popular in the 70s thanks to the publication of The Limits to Growth [3], a book that used an SD model to expose the futility of the idea of infinite growth on a finite planet. Today, the most comprehensive reference about system dynamics is the book Business Dynamics by MIT’s Prof. John Sterman \cite{sterman}.

System dynamics can model both the technical and social aspects of the complex systems established with the adoption of the bitcoin and other cryptocurrencies. Therefore, it is a perfect approach to study the economic dynamics of these new forms of money thanks to its ability to explain emergent systemic phenomena in terms of interactions between factors related to both human behavior and (technical) system architecture.

Surprisingly, little attention has been paid so far to the bitcoin and other cryptocurrencies by researchers in system dynamics. Only two studies so far applied quantitative SD models to cryptocurrencies, presenting an Accounting System Dynamics macroeconomic perspective on peer-to-peer monetary systems \cite{yamaguchi1, yamaguchi2}. Neither of these studies, however, explicitly models or builds upon the unique features of the bitcoin cryptocurrency system architecture.

This paper intends to show how SD can be both a powerful and accessible tool to model the architecture of the blockchain and related market behaviors, that is the technical and social aspects of the bitcoin as a complex system. The power of SD in this context lies in the capability to integrate these aspects in one model, and its accessibility in the visual representation of the model, in terms of stocks, flows, and feedback loops, which make SD models intelligible to a large audience.

\newpage

\section{The Hypothesis}

The model of this paper addresses: what does drive the growth of bitcoin mining power (i.e., the network hash rate)? The hypothesis is that miners mine blocks because they profit. In other words, because 

\[ \textrm{Mining Revenues} \ge \textrm{Mining Cost} \] 

where 

\[ \textrm{Mining Revenues} = (\textrm{Subsidy} + \textrm{Transaction Fees}) \cdot \textrm{BTC Price} \]

\begin{figure}[h]

\begin{subfigure}{0.5\textwidth}
\includegraphics[width=0.9\linewidth]{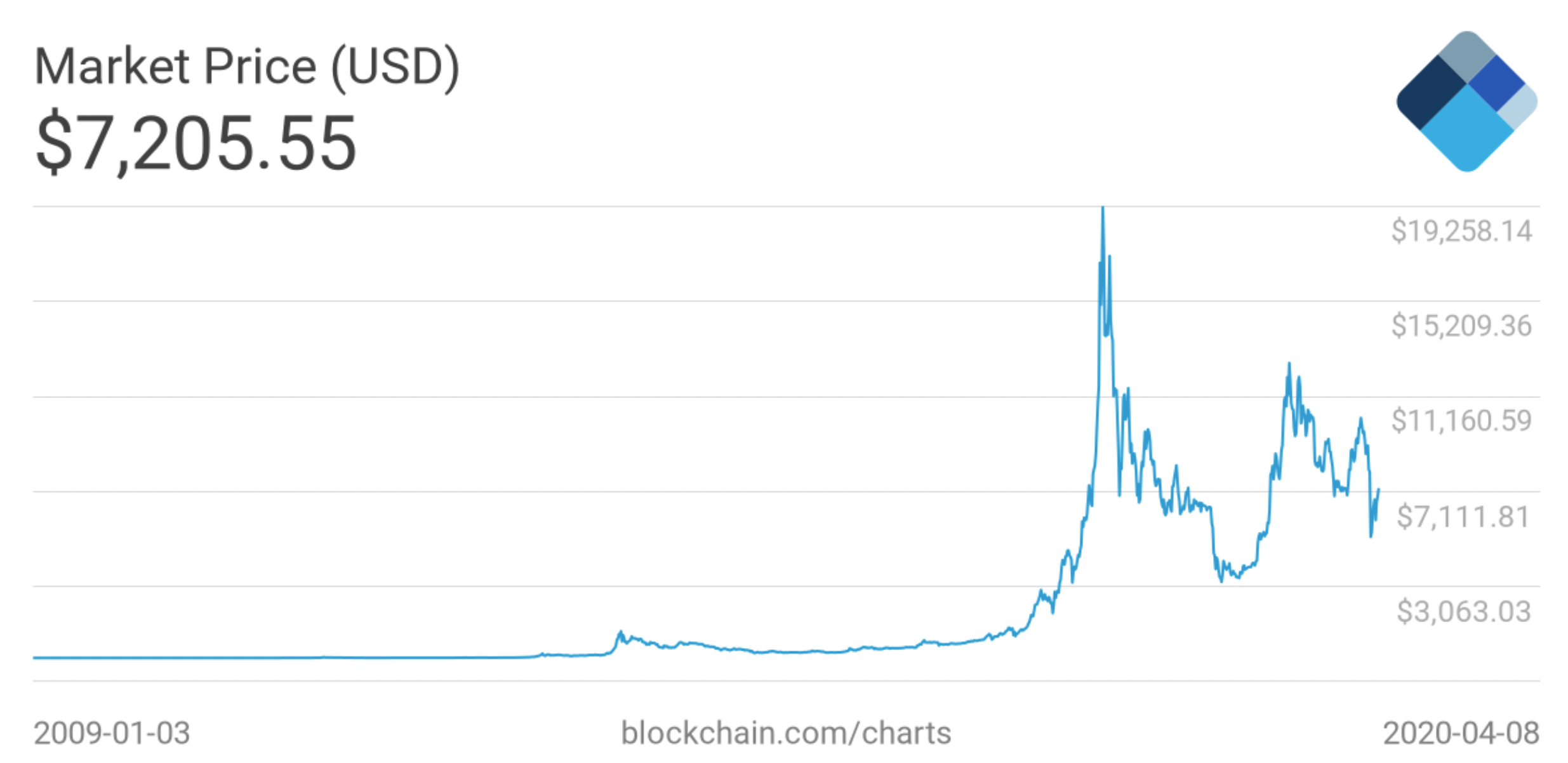} 
\caption{Bitcoin price (in USD/BTC)}
\label{fig:subim3}
\end{subfigure}
\begin{subfigure}{0.5\textwidth}
\includegraphics[width=0.9\linewidth]{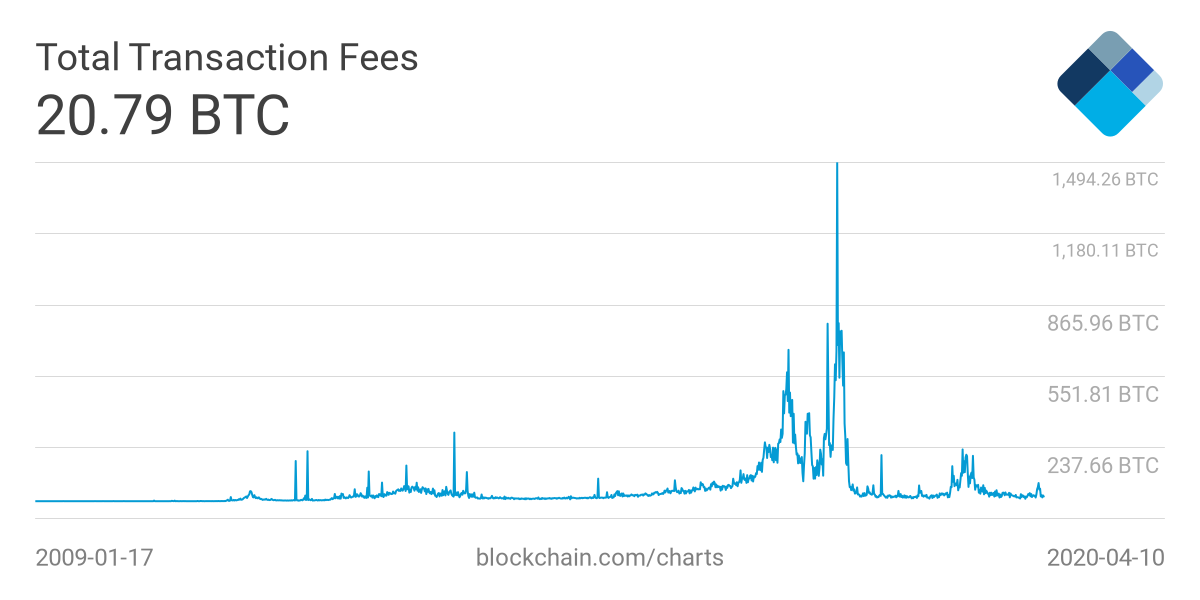}
\caption{Transaction fees (in BTC)}
\label{fig:subim4}
\end{subfigure}

\caption{Historical time series used as exogenous variable in the model. Source of data and images: blockchain.com}
\label{fig:image1}
\end{figure}

Note that both sides of the first equation are in U.S. Dollars (USD) and refer to the reward and cost of mining in monetary terms at the exchange rate between Bitcoin (BTC) and USD at the time the block is mined. For reference, the historical evolution of bitcoin price and transaction fees from the beginning of the bitcoin to the date of writing is given (Fig. \ref{fig:image1}).

In the simplified model of this paper, the operational cost of mining (OPEX) is approximated just with the cost of the energy used in the mining process, whereas other operational costs, as well as the original investment to purchase and setup the mining hardware (CAPEX), are neglected. Finally, it is assumed that energy is traded in a legal tender currency, USD, and therefore miners have to sell bitcoin to pay the electricity bill.

This dynamic hypothesis further assumes an efficient market with perfect information. As long as the mining profit is positive, more hashing power is added to the network until the marginal cost of adding more hashing power exceeds its expected return. Conversely, when the mining profit becomes negative, hash rate is subtracted from the network. The increase and decrease of hash rate happens with a certain time constant, represented with a delay factor (i.e., the adjustment is not instantaneous). In a first approximation, this delay is assumed to be the same for both the increase and decrease of hashing rate, although the latter is likely to be smaller than  the former.

This hypothesis is a coarse simplification of reality, and therefore can only provide a partial answer to the question of what drives the bitcoin network hash rate. Specifically, it assumes that all miners work for the immediate profit of mining, and it does not consider that some (most?) miners may be willing to invest even when the current revenues of mining are negative in the expectation of the future returns from rising BTC price. Nevertheless, it is plausible that the perspective of a present-time gain, at any time, is the dominant mechanism underlying the decision of a miner whether or not to invest in additional mining hardware, or mine bitcoin instead of another more rewarding cryptocurrency. 

Furthermore, the same hypothesis is applied throughout the entire life of the bitcoin, from the early days (January 2009) till now (April 2020). It is obvious that the dynamics driving the mining of the bitcoin in the early days were not the same of today. At the beginning, a true market barely existed, and mining was performed by enthusiastic individuals with modest investment in hardware infrastructure. Today, there is a market with ‘perfect information’ that is dominated by corporations with a stake in mining infrastructure worth any million of USD. Hence, the model is expected to correlated with historic data better in the late years than in the early years.

\section{The Model}

The model of this paper has been created with Vensim (Ventana Systems, Inc.), and it is available at: github.com/davidelasi/bitcoin. The model is organized in two views: Coin Creation  and Hash Rate Adjustment. Certain variables of the system are treated as exogenous variables, meaning that neither their value is determined by the state of other variables of the system, nor they take part in feedback loops.

The following historical time series are exogenous (inputs) to the model:

\begin{itemize}
\item Bitcoin price [in USD/BTC]
\item Revenues from transaction fees [in BTC/Day]
\item Energy efficiency of state of the art mining hardware [in GH/J]
\end{itemize}
The latter is based on our synthesis of a multitude of internet sources that are too many to be mentioned here (from Wikipedia to old pages of other websites and forums about the energy efficiency of mining hardware).

The model correlation is performed by comparing selected model outputs with the historical time series:

\begin{itemize}
\item Height of the Blockchain [in blocks]
\item BTC in circulation [in BTC]
\item Hash Rate [in GH/s]
\end{itemize}
The first two are used only as a model sanity check, whereas the last one is used for model correlation. The correlation of the model has been performed both manually and using the optimization capabilities of Vensim DSS.

The model structure is discussed in detail in the following sections. The following conventions are used: variable names are capitalized, and constant names are all capital. Exogenous variables are highlighted in gray in the figures. Bitcoin is abbreviated as BTC.

\subsection{Coins Creation}

\begin{figure} [h]
    \centering
    \includegraphics[width=4in]{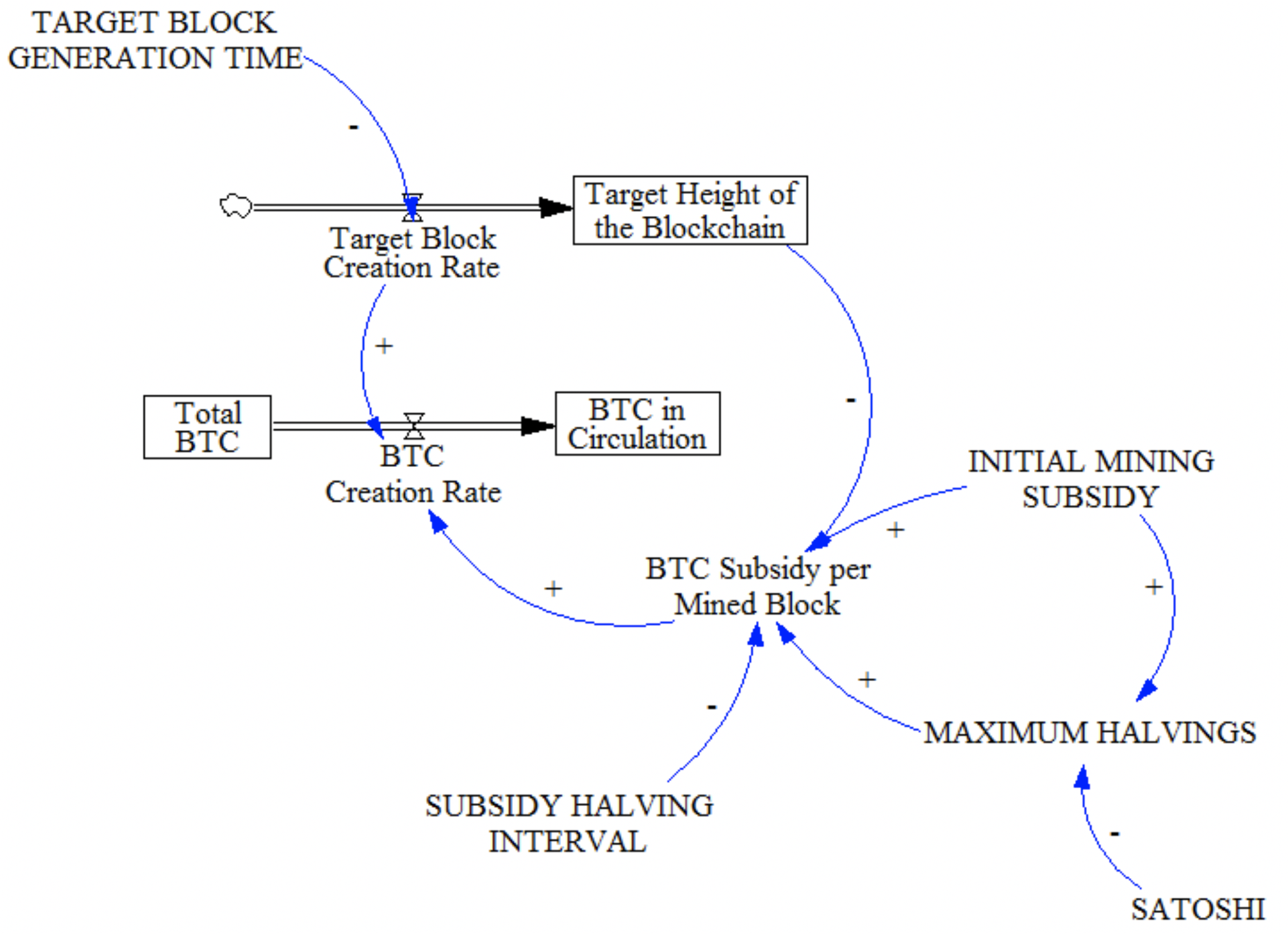}
    \caption{Coin Creation model view.}
    \label{fig:coin_creation}
\end{figure}

The creation of bitcoins is modelled as shown in Fig. \ref{fig:coin_creation}. This diagram represents the mechanism of controlled supply devised by Satoshi Nakamoto. 

There is a finite stock of 21,000,000 Total BTC, whose content flows into a second stock of BTC in Circulation with a certain creation rate. This creation rate is given by the product of two factors: how many blocks are mined per day, and how much subsidy (in BTC) is granted for each mined block. The former is assumed to correspond to the Target Block Creation Rate of 1 block every 10 min (144 blocks / day), and the latter is a geometric series following the process that is hard-coded in the bitcoin software: an initial subsidy of 50 BTC per mined block, that is halved every 210,00 blocks until the value of the subsidy reaches the minimum bitcoin-transaction unit of 0.00000001 BTC, commonly known as ‘1 Satoshi’. 

This part of the model does not include any dynamic loop. No feedback is present, and with respect of the dynamic hypothesis current under investigation, the only connections to the higher layer of the model are through the variables BTC Subsidy per Mined Block and Target Block Creation Rate, which are used to compute the Revenues from Subsidy.

\newpage

\subsection{Hash Rate Adjustment}

\begin{figure} [h]
    \centering
    \includegraphics[width=5in]{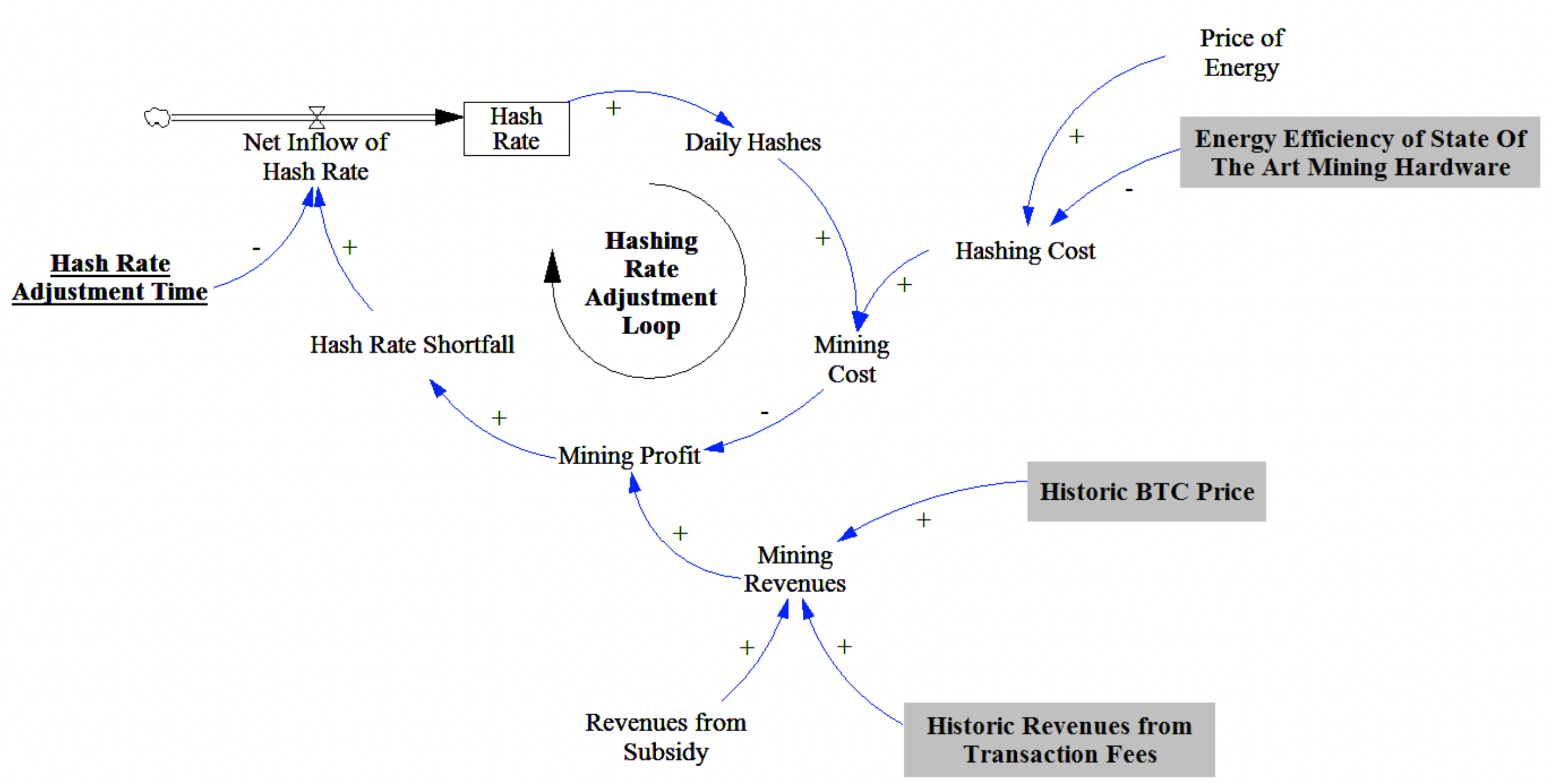}
    \caption{Hash Rate Adjustment model view. Exogenous variables in gray.}
    \label{fig:hash_rate_adjustment}
\end{figure}

The addition of hash rate to the bitcoin network is modeled as shown in Fig. \ref{fig:hash_rate_adjustment}. The negative feedback loop is the manifestation of the effect of the efficient-market hypothesis set forth earlier.

The network Hash Rate (GH/s) is a stock with a Net Inflow of Hash Rate representing the amount of computational power that is daily added or subtracted from the network.  The Hash Rate at time zero is set to 0.007 GH/s, according to some estimation of the original computational power available to Satoshi Nakamoto in the early days of bitcoin \cite{satoshi-hash}. From the Hash Rate, the number of Daily Hashes (GH/day) computed by the network is calculated.

The Mining Cost is determined by the Hashing Cost, expressed in \$/GH and given by the quotient between the Price of Energy (\$/J) and the Energy Efficiency of State of the Art Mining Hardware (GH/J). The Price of Energy is assumed to be constant over the studied time period at 0.05 \$/kWh.  This is a simplification that may not be fully representative of the average price of energy used by miners in the early days, although it seems to be aligned with the present (according to \cite{coinshare-research}, ~0.04 \$/kWh).

\begin{figure} [h]
    \centering
    \includegraphics[width=3in]{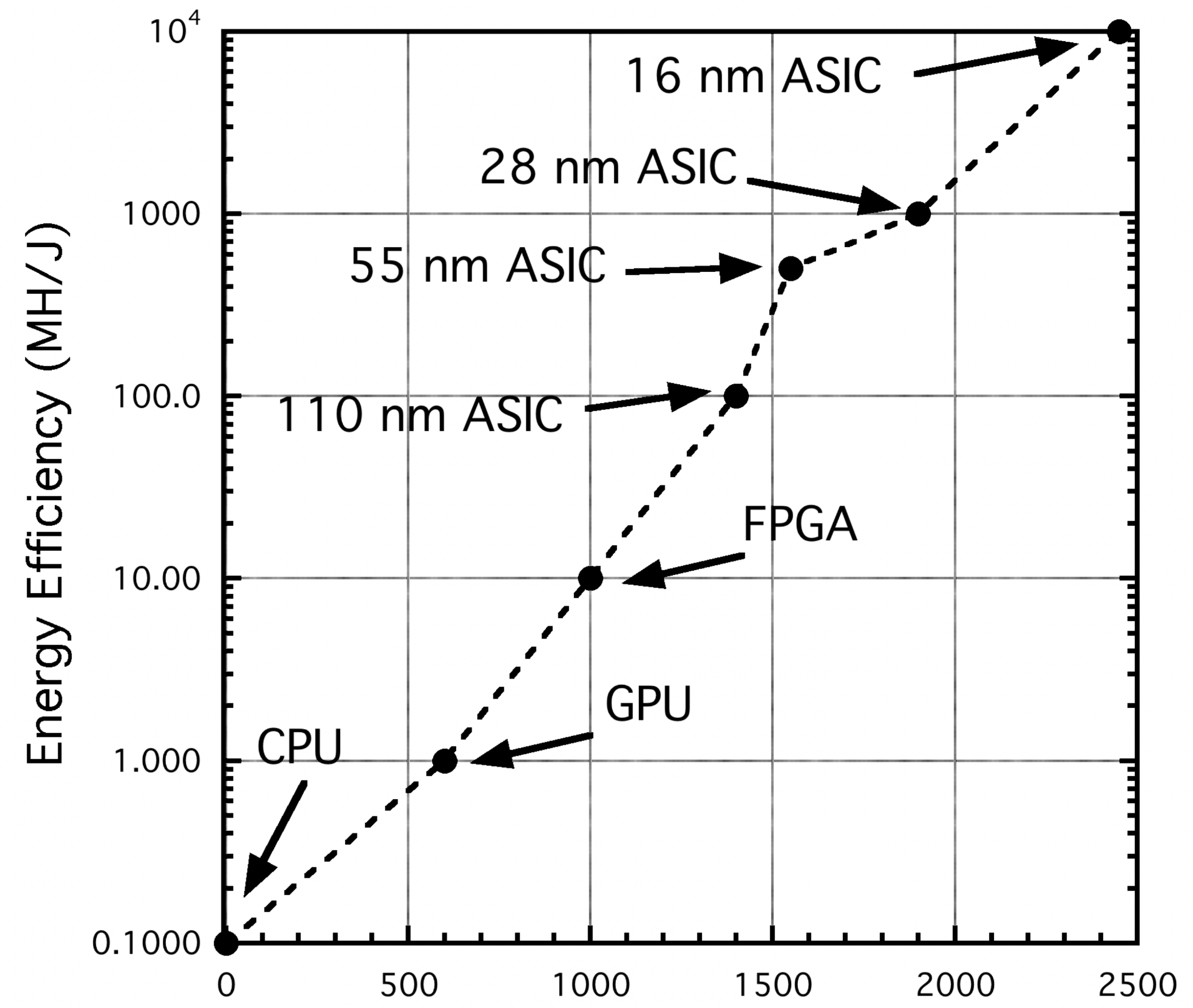}
    \caption{Energy efficiency evolution of state-of-the-art mining hardware.}
    \label{fig:miners_technology}
\end{figure}

The Energy Efficiency of State of the Art Mining Hardware is modeled as a constant level in different epochs of mining. Seven periods are identified, from the times of the CPU to the most-recent ASIC technology (Table \ref{table:1} and Fig. \ref{fig:miners_technology}).  At the end of each period, the variable assumes the value of the next period, as if the whole network would immediately switch to the new technology at once. This is not realistic and creates discontinuities in the time series, as the difference of this variable between consecutive period is often of the order of 10x. The model could be easily improved by adding further elements to model the typical S-shaped (logistic function) pattern of new technology's diffusion.

\begin{table}[h!]
\begin{center}
\begin{tabular}{ |c|c|c|c| } 
 \hline
 Technology & From (date) & Period (days) & Energy efficiency (MH/J) \\
 \hline
 CPU & Mar 2009 & 0–600 & 0.1 \\
 GPU & Oct 2010 & 600–1,000 & 1 \\
 FPGA & Dec 2011 & 1,000–1,400 & 10 \\
 ASIC (110 nm) & Jan 2013 & 1,400–1,550 & 100 \\
 ASIC (55 nm) & Jun 2013 & 1,550–1,900 & 500 \\
 ASIC (28 nm) & Jun 2014 & 1,900–2,450 & 1,000 \\
 ASIC (16 nm) & Dec 2015 & 2,450–Now & 10,000 \\
 \hline
\end{tabular}
\caption{Evolution of state-of-the-art mining hardware}
\label{table:1}
\end{center}
\end{table}

The Mining Revenues are calculated by summing the Revenues from Subsidy and from transaction fees, and converting them into \$. This calculation uses a mix of endogenous and exogenous factors: the subsidy is calculated based on the variables modelled in the ‘Coins Creation’ view described in the previous section, whereas the transaction fees and the BTC price are extracted from historic data series mentioned earlier. 

Subtracting the cost from the revenues of mining, the Mining Profit is calculated. This determines the amount of Hash Rate Shortfall, that is how much hashing power could be added before the marginal cost of adding more will exceed the benefit (a situation of zero-profit that is the goal of the negative feedback loop):

\[\textrm{Hash Rate Shortfall} = \frac{\textrm{Mining Profit}}{\textrm{Cost of Energy} \cdot \textrm{Energy Efficiency of Mining Hardware}} \]

\newpage

In other words, the profit that the network makes every day (\$/day, expressed in \$/s) is converted into an amount of energy (J/s) that the network could afford to add before reaching a zero-profit situation. This amount of energy, in turn, corresponds to a certain amount of hashing power (GH/s) calculated through the Energy Efficiency of State of the Art Mining Hardware that is available at any given time (GH/J).

At last, the key factor involved in this loop is introduced: the Hash Rate Adjustment Time. This accounts for the fact that the system does not react instantaneously to a shortfall in hash rate, for instance, because the supply chain of mining hardware is (notoriously) limited in its throughput. This parameter is the only one that is acted upon for the correlation of the model results by comparing the calculated hash rate with the historical time series. 

\section{Results and Discussion}

The results and discussion are divided into three sections about model correlation, insights about the past, and speculations about the future. Table \ref{table:2} maps the number of days of the simulation to calendar dates and relevant events, to help with the interpretation of the results.

\begin{table}[h!]
\begin{center}
\begin{tabular}{ |c|c|c| } 
 \hline
 Time (day) & Date & Comment \\
 \hline
 0 & 3 Jan 2009 & Mining of the genesis block \\
 500 & May 2010 & Introduction of GPU miners \\
 1,000 & Sep 2011 & Introduction of FPGA miners \\
 1,500 & Feb 2013 & Introduction of 1st gen. ASIC \\
 2,500 & Nov 2015 & Introduction of 4th gen. ASIC \\
 3,300 & Dec 2017 & Bitcoin grazes the 20,000 \$/BTC price mark \\
 3,700 & Feb 2019 & – \\
 4,100 & Mar 2020 & Latest update of this model \\
 \hline
\end{tabular}
\caption{Evolution of state-of-the-art mining hardware}
\label{table:2}
\end{center}
\end{table}

\subsection{Model Sanity Check and Correlation}

\begin{figure} [h]
    \centering
    \includegraphics[width=2.1in]{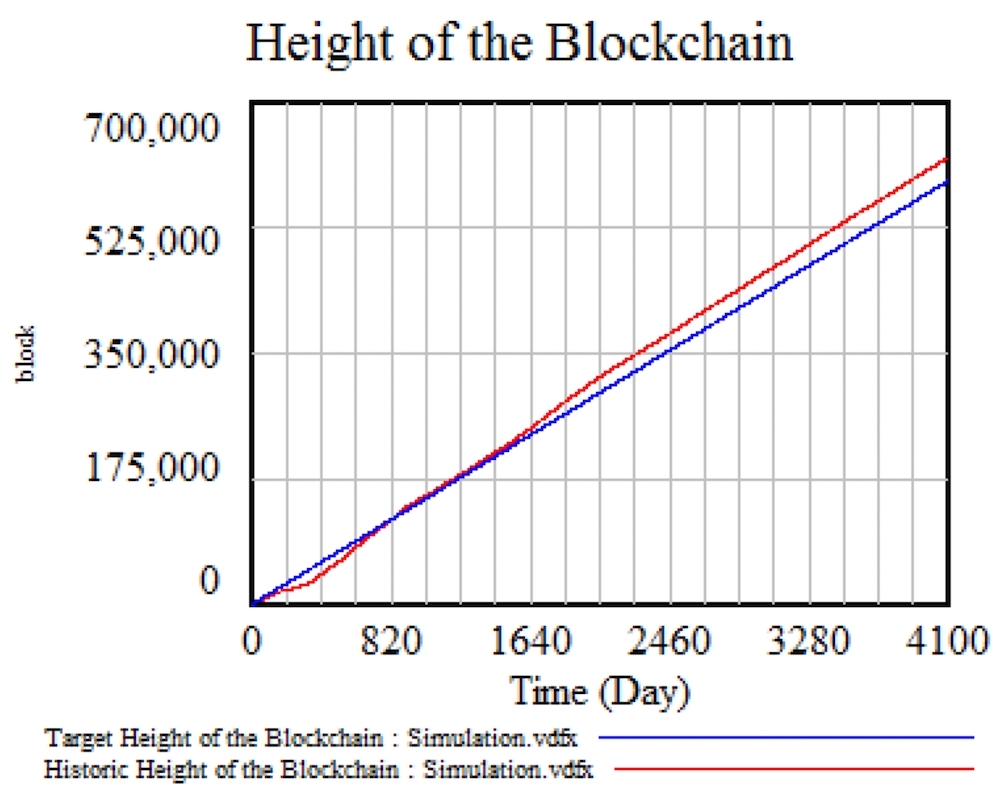}
    \caption{Evolution of the height of the blockchain over time (model vs reality).}
    \label{fig:blockchain-height}
\end{figure}

A sanity check of the model is performed by comparing the height of the blockchain calculated by the model with the actual historical time series (Fig. \ref{fig:blockchain-height}). The comparison is satisfactory, showing that the Coin Creation part of the model works as expected.

The correlation of the model with historical data series has been performed by tuning the value of the Hash Rate Adjustment Time manually, and through the automated optimization routines, until reaching a satisfactory matching between the calculated and the historical time series of the hash rate.

\begin{figure}[h]

\begin{subfigure}{0.5\textheight}
\includegraphics[width=12cm]{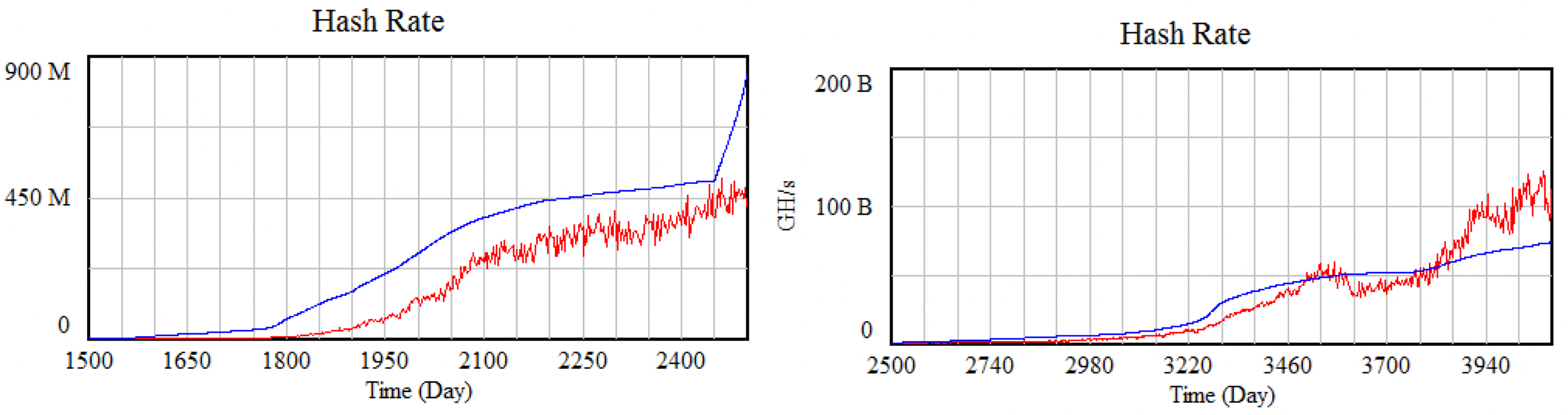} 
\caption{Correlation with one delay value.}
\label{fig:subim1}
\end{subfigure}
\begin{subfigure}{0.5\textheight}
\includegraphics[width=12cm]{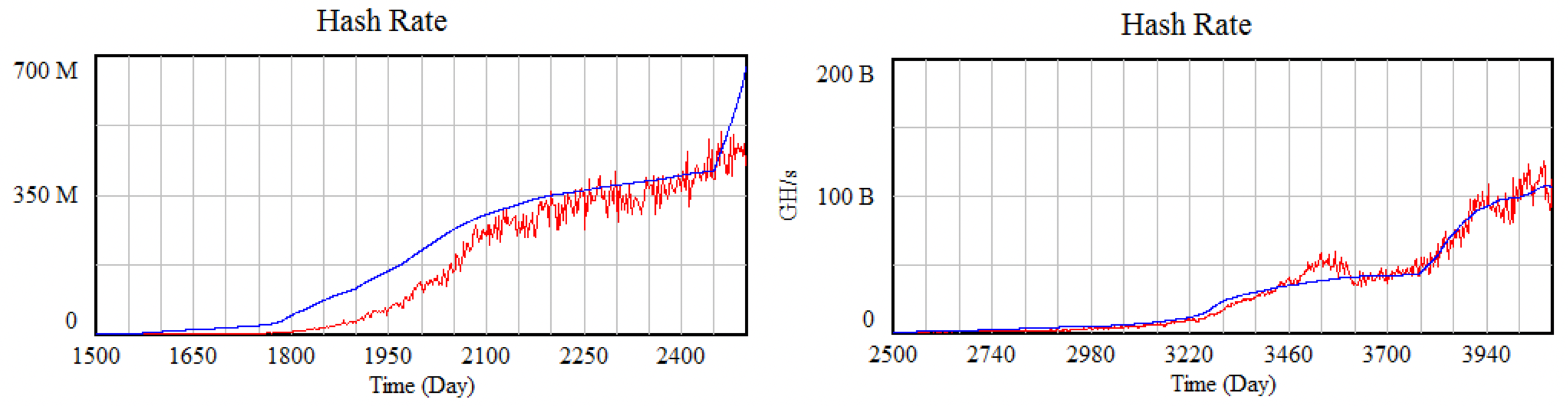}
\caption{Correlation with two delay values.}
\label{fig:subim2}
\end{subfigure}

\caption{Model correlation with Hash Rate Adjustment Time: top row 1112 days; bottom row, 1482 days (time 0 – 3,777) and 264 days (time 3,777 – 4,100).}
\label{fig:hash-rate-correlation}
\end{figure}

By using a single value for the Hash Rate Adjustment Time throughout the whole time series, an optimal result can be found at 1112 days, but a poor correlation is obtained, particularly in the last three years (Fig. \ref{fig:hash-rate-correlation}a). The situation can be significantly improved by allowing for the possibility that the delay time of the feedback loop changes at some point (i.e., IF time < value THEN use delay time 1 ELSE use delay time 2). The optimal fit occurs for a delay time of 1482 days from time 0 to 3,777 and only 264 days from time 3,777 to 4,100 (Fig. \ref{fig:hash-rate-correlation}b). The rest of the results in this paper are relative to this second scenario.

\newpage

In general, the correlation can be considered good enough to gain some insights from this model, especially considering its simplicity, with the following provisos:

\begin{itemize}
    \item The Hashing Rate Adjustment Loop does not capture well the dynamics of the system in the early days (0-1,500 days) dominated by CPU, GPU, and FGPA miners. Model improvements are needed to use it for this period. For example, the diffusion of new technologies could be modeled to allow for a gradual implementation of the newer generation of mining hardware, particularly in the early days, rather than as a step function. This could reduce the tendency of the model to overestimate the hash rate in the early days, and it would also avoid discontinuities in the hash rate as for time 2,450 in Fig. \ref{fig:hash-rate-correlation}. Also, a better estimate of the Price of Energy of the miners in the early days, which as certainly higher on average than 0.05\$/kWh, would help.
    \item The model reasonably captures the main dynamics of the system from time 1,500 on, which is in the epoch of ASIC miners; however, some dynamics are not represented by the model (e.g., the peak in Hash Rate around time ~3,600). To capture these ‘modes’, one would need to envisage and implement additional feedback loops working together with the Hashing Rate Adjustment Loop.
    \item Using two different adjustment times with an IF THEN condition is a ‘quick and dirty’ trick, which is generally considered ‘bad practice’ in a system dynamics model. The need to use such a ‘trick’ is symptomatic of the fact that the Hash Rate Adjustment Time might be, itself, part of other feedback loops that are currently not modeled, which would justify the change of this variable with time. For example,  the delay’s decrease could be due to the reinvestment of mining profits in new technology or the enhancement of the supply chain of existing technologies. 
\end{itemize}

In summary, there are several opportunities to extend and improve this model by identifying and implementing these additional loops. Yet, this simple model is sufficiently complex to gain relevant insights about the system while avoiding the risk of over-fitting the data.

\newpage

\subsection{Insights About The Past}

\begin{figure} [h]
    \centering
    \includegraphics[width=3in]{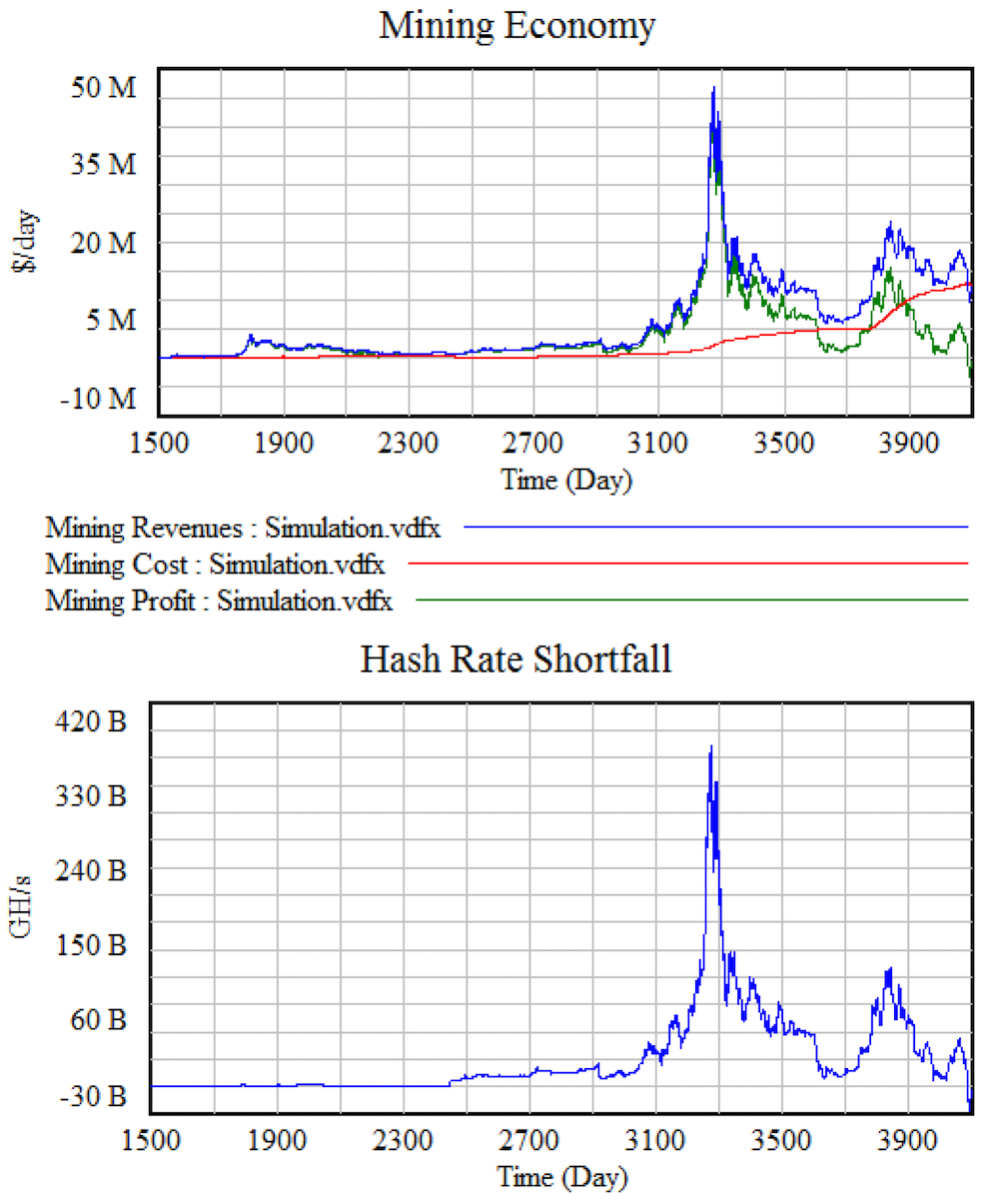}
    \caption{The Hash Rate shortfall that drives the negative feedback loop, and the mining economy in terms of revenues, cost, and profit (results for past time).}
    \label{fig:mining-past}
\end{figure}

Having established a reasonable correlation between the model and the reality, the analysis of the simulation results gives further insights on the past evolution of bitcoin mining and the related market dynamics:

\begin{itemize}
    \item The long delay time of the feedback loop until ~ ten years (day 3,777) from the genesis block, means that there was much more potential demand than supply of mining hardware. In other words, even if mining was a great deal financially speaking, procuring as much hashing power as it would financially make sense from a pure present-day profit consideration, was never possible. The long delay time of hash rate addition is also in line with the lived experience that mining hardware has always been a (relatively) scarce resource, and further invalidates some efficient-market hypothesis models which would overestimate hash rate growth.  
    \item The jump to a much lower delay time of the feedback loop that occurred around mid-2019, from 1,482 to 264 days, is also aligned with some factual evidence. In particular, in 2019, new mining hardware with ~5x the hashing power per unit sold compared to the previous state-of-the-art hardware became available \cite{coinshare-research}. This ratio matches quite well with the ratio between the two delays time, which is 5.6, and it would be compatible with the same supply chain delivering approximately the same ‘flow’ of miners (in units) but a ~5x bigger flow of mining power (in hash rate).
    \item Until late 2019, there was always a positive shortfall of hashing power in the network, with the biggest spike matching the time of highest bitcoin price and transaction fees. Compare Fig. \ref{fig:mining-past} with Fig. \ref{fig:image1}. The profit from mining, expressed in the simplistic terms of this paper as same-day revenues minus operational costs accounting for the cost of energy, was always positive, until the last few months. The current price of ~7,000 \$/BTC is likely to be close to the marginal cost of production of a bitcoin in the present conditions. The latter point, in particular, calls for some thinking about the future and the reaction of the system to the forthcoming halving event, in about a month from the time of writing. 
\end{itemize}

\newpage

\subsection{Speculations About The Future}

Using a model correlated with past data to attempt to predict the future is always a tricky endeavor. This is true for this model as well for two reasons. First, because it is a simple model with only one feedback loop it may not capture all the dynamics of the system. Second, because the model correlation is performed with past data that were almost always relative to a growing network hash rate, and there is no guarantee that the same dynamics work when the market would turn in a different direction. In fact, the limited ability of the model to represent the ‘mode’ of the peak at 3,600 days (to the right in Fig. \ref{fig:hash-rate-correlation}) signals that different dynamics might apply to a decrease of the network hash rate. 

With this necessary premise, the temptation to apply the model to a hypothetical future is irresistible. Therefore, we have exercised the model after a hypothetical future scenario where the price of bitcoin and the revenues from transaction fees remain at an average level close to the recent time’s. This scenario is interesting because the current profit from mining, net of the cost of energy, is nearly zero (Fig. 8), meaning that the hashing power may have reached a status of equilibrium at the present level of revenues from the subsidy and the transaction fees. Therefore, it is interesting to see that this equilibrium might be perturbed by the forthcoming halving of the subsidy.

Specifically, we have extended the model operation from day 4,100 to 7,500 with the following assumption:

\begin{itemize}
    \item The Future BTC Price fluctuates around an average level of 7,300 \$/BTC. This is represented with a pink-noise time series (std deviation 500 \$/BTC and correlation time of 28 days). 
    \item The Future Revenues from Transaction Fees fluctuate around an average level of 30 BTC/day. This is also represented with a pink-noise time series (std deviation 5 BTC/day and correlation time of 28 days).
\end{itemize}

All these assumptions are of course arbitrary, and one could play with any different set of assumption.

The Energy Efficiency of State of the Art Mining Hardware is kept constant at today’s value, which is probably not too far off from reality in the short term, considering that orders of magnitude improvements of the present ASIC technology are unlikely to happen (although in the long-term quantum computing might be a game changer). 
While bitcoin mining spurs further technological advancement along the lines of Moore’s law, we don’t expect massive technological changes to affect this system dynamics model.  To put it another way, the constraint of reduction of coinbase reward is expected to dominate over technological leaps in hashing efficiency.

\begin{figure} [h]
    \centering
    \includegraphics[width=3in]{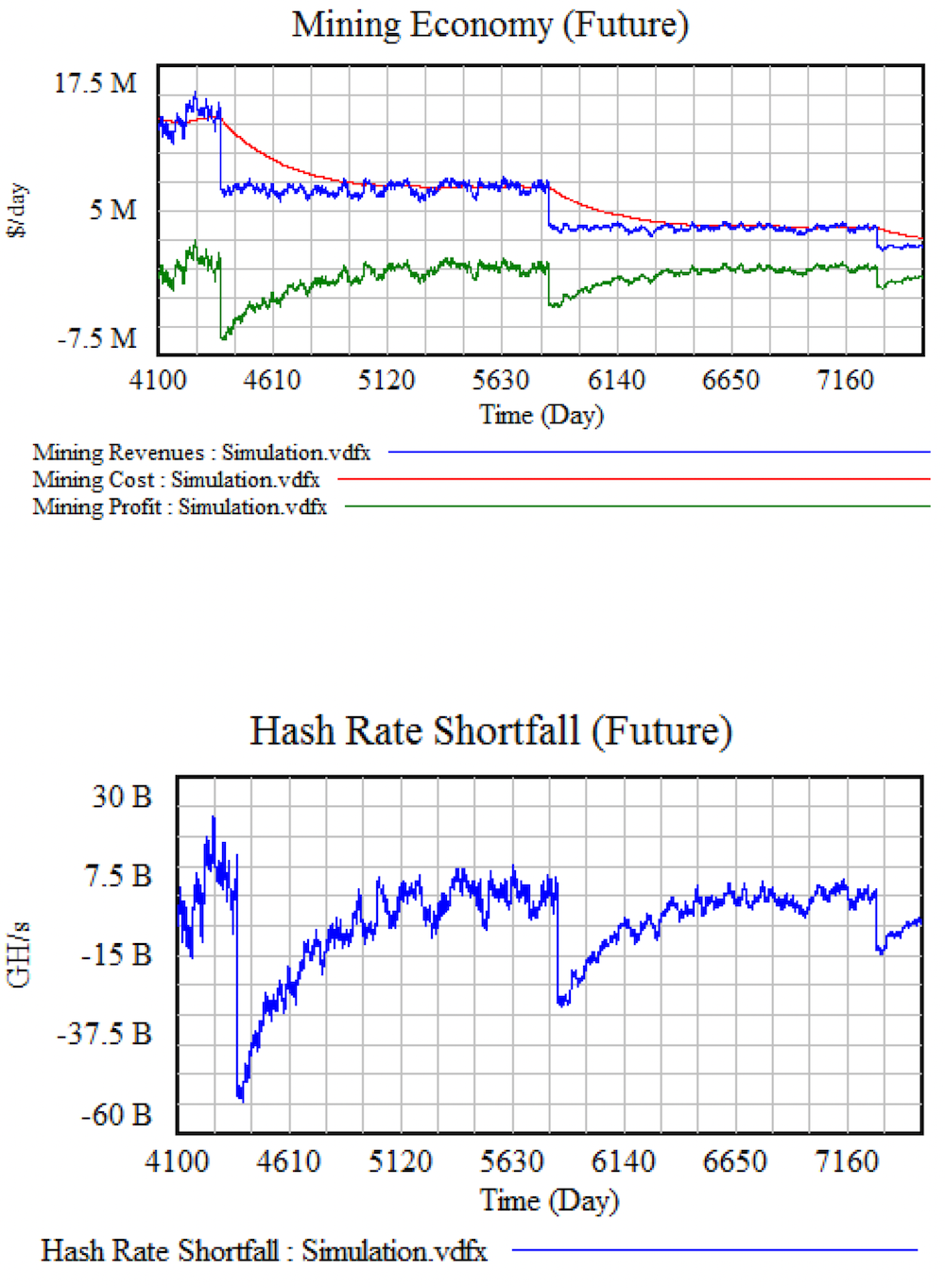}
    \caption{The Hash Rate shortfall that drives the negative feedback loop, and the mining economy in terms of revenues, cost, and profit (future projection, see the text for details about the assumed revenues from subsidy and transactions).}
    \label{fig:mining-future}
\end{figure}

Fig. \ref{fig:mining-future} shows the projected future evolution of the revenues, costs, and profits from mining, and the evolution of the Hash Rate Shortfall determines the operation of the negative feedback loop of the model. The results show a quick switch to negative-profit for block mining at the next two halving events, and particularly at the next one, when the mining subsidy will be cut in half for the third time, from 12.5 to 6.25 BTC per block. Correspondingly, the Hash Rate Shortfall becomes negative, to indicate that the network has an excess of hash rate, which makes the cost of mining higher than the revenues that it generates. In other words, should the (dollar) revenues from mining remain at today’s level, mining will suddenly start to become unprofitable an unprofitable business until the difficulty goes down because of a reduction of the network hash rate. 

The feedback loop of the model, based on the efficient-market hypothesis set forth at the beginning of the paper, operates by reducing the hash power of the network until the net profit from mining will tend again to zero. The difference from the past is that, for the first time, the hash rate would tend to its equilibrium state from the bottom instead of the top. 

In the current model, the adjustment time of the net inflow hash rate is the same, regardless of the sign (i.e., addition or removal of hash rate). However, it is likely that the removal of hash rate will take place with a much shorter adjustment time, because it is much easier to switch off a mining hardware or use it for another purpose (e.g., switch to another cryptocurrency) than to procure and set it up. This could be implemented by considering different inflow and outflow of hash rate from the Hash Rate stock (currently modeled as a single net flow, see Fig. \ref{fig:hash_rate_adjustment}). This effect is not considered in the model, but it can be added in the future if the network hash rate will indeed start to drop. Hence there will be data to correlate this ‘mode’ of the system (in fact, there may already be some data to try to do so using the peak of hash rate at 3,600 days, to the right in Fig. \ref{fig:hash-rate-correlation}).

\begin{figure} [h]
    \centering
    \includegraphics[width=3.5in]{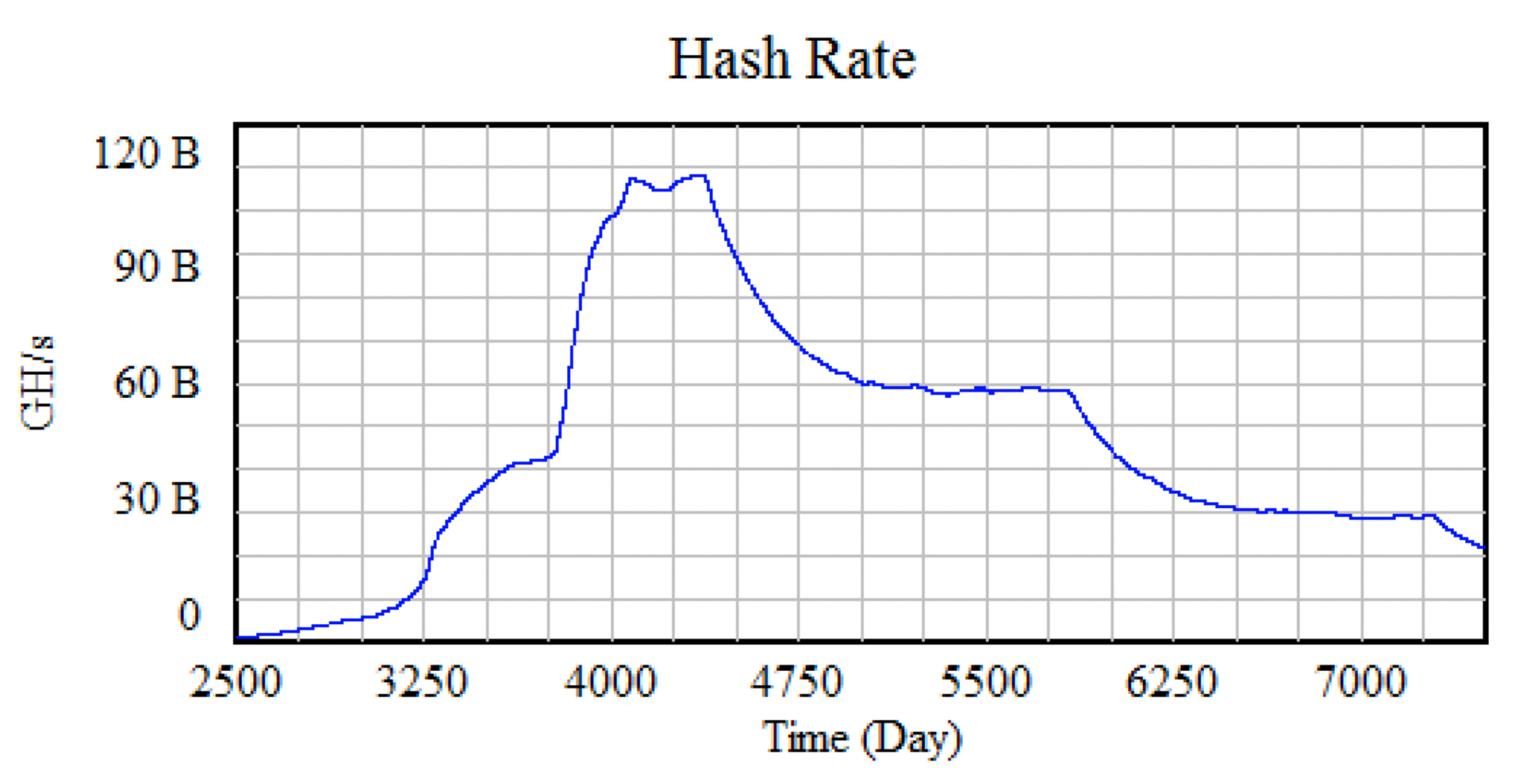}
    \caption{Peak hash (present time = 4100).}
    \label{fig:peak-hash}
\end{figure}

This brings us to the last final question of this work: has the bitcoin reached the ‘peak hash’? That is, has the bitcoin reached the situation where the cost of mining is higher than the revenues from mining? Possibly, as the projection of the future evolution of the hash rate in Fig. \ref{fig:peak-hash} shows. Under the future (halved) value of the subsidy, the equilibrium hash rate of the system might be closer to ~60B GH/s than to the current ~110 GH/s. 

If that will be the case, a most interesting question will be: where will all that excess hashing power be invested, and how this might change the landscape of the bitcoin and other cryptocurrencies? Could the bitcoin and other networks’ hash rate experience higher volatility of hash rate after the next halving event? May such a staggeringly large amount of hashing power that could potentially flee from the bitcoin network in the long-term be a threat to the security of the network (or other SHA256 secured coins such as BCH and BSV), if ever it would suddenly come back? 

Of course, a number of events may allow for a smooth transition between the current and the next reward era, such as significant improvements in the energy efficiency of the mining hardware, access to yet cheaper energy, or substantial inflation of bitcoin price and transaction fees. Some of these changes (e.g., increase in block size) and trends (e.g., increase in transaction fees) have been already happening in the past. But one thing seems true: the effect of the forthcoming halving might be different and have deeper consequences than the previous, so it is worth making sure that the bitcoin, as a dynamical system, is prepared for this transition.

Finally, it is worth considering what could be an alternative model of coin supply that would not have incurred this (hypothetical) peak of hash rate. In 2014, for instance, a cryptocurrency based on a continuous logarithmic coin release mechanism was released: the Woodcoin [10], which replaces bitcoin’s geometric series reduction of coinbase rewards with a harmonic series. Such a mechanism, among others, ensures a slower overall rate of coinbase reduction, changing the system dynamics especially in the longer term. While the “halving” system of a geometric series reduction of coinbase chosen by Satoshi has been popular among other public currencies (e.g. XMR, LTC), a System Dynamics model could tell us more about how other choices of currency issuance will affect the hash rate and future security of the system.  

\newpage

\section{Conclusions}

We have shown that aspects of bitcoin mining can be modeled as a dynamical system using system dynamics. Starting with a dynamical hypothesis based on an efficient-markets hypothesis applied to the mining of blocks, we have shown how the recent evolution of the bitcoin network has rate can be explained by a negative feedback dynamic loop that zeroes on mining profit with a delay time. 

By extending the simulation into the future to cover the next two reward eras, under the assumption of constant bitcoin price and revenues from transaction fees approximately at the level of April 2020, we have shown how there is a possibility that the next halving event lead to a transition to an unprofitable mining regime with an excess of network hash rate. This may imply that we might be at the ‘peak hash’, and experience a continuous decline of the hash rate (and difficulty) in the coming years. 

Finally, this model shows that system dynamics methods and tools can be effective to model the bitcoin and could be applied to other existing or new proposed cryptocurrencies as well to understand the behavior of the complex sociotechnical systems that are created from the application of blockchain technology to electronic money and other applications enabled by shared ledger.  It is worth noting the relative success that the model enjoys with historical data, especially considering that there is no way to verify how many people are mining nor what hardware is available to them.

\subsection*{Future Work}

Despite a decade of practice, the theory behind public proof-of-work ledgers remains in its infancy.  There are many additional tweaks possible to the system dynamics we have presented here, including more complex models of mining efficiencies, other currency release curves (other coins), and modeling the effect of market shocks.  The initial applicability of the model leaves us expecting much more from future work using system dynamics to understand the evolution of the public proof-of-work consensus network.

\subsection*{Acknowledgments}

One of the authors, Davide Lasi, is extremely grateful to Prof. Jim Lyneis, lecturer of system dynamics in the MIT \textit{System Design \& Management (SDM)} program. His teachings of system dynamics and his generous dedication of time to brainstorm about system dynamics and the bitcoin were instrumental for laying the ground for this work in the summer of 2017, when this model was conceived. Prof. Lyneis has not reviewed the final implementation of this model; therefore, any mistake is only attributable to the authors.

\end{document}